\documentclass[10pt,letterpaper]{article}
\usepackage[top=0.85in,left=1.25in,right=1.25in,footskip=0.1in]{geometry}
\usepackage{tabularx}
\usepackage[utf8]{inputenc}
\usepackage{url}
\usepackage{todonotes}
\usepackage{tabularx}
\usepackage{array}
\usepackage{graphicx} 
\usepackage{multirow}
\usepackage{hhline,booktabs,amsmath}
\usepackage{subcaption}
\usepackage{cite}

\usepackage{nameref,hyperref}

\usepackage[right]{lineno}

\usepackage{microtype}
\DisableLigatures[f]{encoding = *, family = * }

\usepackage{caption}

\usepackage{url}
\usepackage{todonotes}
\usepackage{tabularx}
\usepackage{array}
\usepackage{graphicx} 
\usepackage{multirow}
\usepackage{hhline,booktabs,amsmath}
\usepackage{subcaption}
\makeatletter
\renewcommand{\@biblabel}[1]{\quad#1.}
\makeatother

\usepackage{lastpage,fancyhdr,graphicx}
\usepackage{epstopdf}
\usepackage{color}

\definecolor{Gray}{gray}{.25}

\usepackage{graphicx}

\usepackage{wrapfig}
\usepackage[pscoord]{eso-pic}
\usepackage[fulladjust]{marginnote}
\reversemarginpar
\begin{document}
\vspace*{0.35in}

\begin{flushleft}
{\Large
\textbf\newline{Onto2Vec: joint vector-based representation of
  biological entities and their ontology-based annotations}
}
\newline
\\
Fatima Zohra Smaili\textsuperscript{1},
Xin Gao\textsuperscript{1,*},
Robert Hoehndorf\textsuperscript{1,*},
\\
\bigskip
\bf{1} Computer, Electrical and Mathematical
  Sciences and Engineering Division, Computational Bioscience Research Center, King Abdullah University of Science and Technology, Thuwal 23955, Saudi Arabia.

\bigskip
* robert.hoehndorf@kaust.edu.sa and xin.gao@kaust.edu.sa.

\end{flushleft}

\section*{Abstract}
{\textbf{Motivation:} Biological knowledge is widely
  represented in the form of ontology-based annotations: ontologies
  describe the phenomena assumed to exist within a domain, and the
  annotations associate a (kind of) biological entity with a set of
  phenomena within the domain. The structure and information contained
  in ontologies and their annotations makes them valuable for
  developing machine learning, data analysis and knowledge extraction
  algorithms; notably, semantic similarity is widely used to identify
  relations between biological entities, and ontology-based
  annotations are frequently used as features in machine learning
  applications.
  \\
  \textbf{Results:} We propose the Onto2Vec method, an approach to
  learn feature vectors for biological entities based on their
  annotations to biomedical ontologies. Our method can be applied to a
  wide range of bioinformatics research problems such as
  similarity-based prediction of interactions between proteins,
  classification of interaction types using supervised learning, or
  clustering.  To evaluate Onto2Vec, we use the Gene Ontology (GO) and
  jointly produce dense vector representations of proteins, the GO
  classes to which they are annotated, and the axioms in GO that
  constrain these classes. First, we demonstrate that
  Onto2Vec-generated feature vectors can significantly improve
  prediction of protein-protein interactions in human and yeast. We
  then illustrate how Onto2Vec representations provide the means for
  constructing data-driven, trainable semantic similarity measures
  that can be used to identify particular relations between
  proteins. Finally, we use an unsupervised clustering approach to
  identify protein families based on their Enzyme Commission 
  numbers. Our results demonstrate that Onto2Vec can generate high
  quality feature vectors from biological entities and
  ontologies. Onto2Vec has the potential to significantly outperform
  the state-of-the-art in several predictive applications in which
  ontologies are involved.\\
  \textbf{Availability:} \url{https://github.com/bio-ontology-research-group/onto2vec}\\
  \textbf{Contact:}
  \href{robert.hoehndorf@kaust.edu.sa}{robert.hoehndorf@kaust.edu.sa and xin.gao@kaust.edu.sa.}
}


\section {Introduction}
Biological knowledge is available across a large number of resources
%
and in several formats. These resources capture different and often
complementary aspects of biological phenomena. Over the years,
researchers have been working on representing this knowledge in a more
structured and formal way by creating biomedical ontologies
\cite{biontologies}. Ontologies provide the means to formally
structure the classes and relations within a domain, and are now
employed by a wide range of biological databases, webservices, and
file formats to provide semantic metadata
\cite{Hoehndorf2015role}. 

Notably, ontologies are used for the annotation of biological entities
such as genomic variants, genes and gene products, or chemicals, to
classify their biological activities and associations
\cite{Smith2007}. An annotation is an association of a biological
entity (or a class of biological entities) and one or more classes
from an ontology, usually together with meta-data about the source and
evidence for the association, the author, etc.~\cite{Hill2008}.

Due to the wide-spread use of ontologies, several methods have been
developed to utilize the information in ontologies for data analysis
\cite{Hoehndorf2015role}. In particular, a wide range of semantic
similarity measures has been developed \cite{Couto2009} and applied to
the similarity-based analysis of ontologies and entities annotated
with them.  Semantic similarity is a measure defined over an ontology,
and can be used to measure the similarity between two or more ontology
classes, sets of classes, or entities annotated with sets of ontology
classes.


Semantic similarity measures can be classified into different types
depending on how annotations (or instances) of ontology classes are
incorporated or weighted, and the type of information from an ontology
that is used to determine similarity \cite{Couto2009,
  Harispe2015}. Most similarity measures treat ontologies as graphs in
which nodes represent classes and edges an axiom involved the
connected classes \cite{Couto2009, Harispe2015}. However, not all the
axioms in an ontology can naturally be represented as graphs
\cite{Smith2005, h20, rodriguez2018inferring}. A possible alternative
may be to consider all axioms in an ontology when computing semantic;
the challenge is to determine how each axiom should contribute to
determine similarity beyond merely considering their syntactic
similarity.

In addition to similarity-based analysis, ontology-based annotations
are frequently used in machine learning approaches. Ontology-based
annotations can be encoded as binary vectors representing whether or
not an entity is associated with a particular class, and the semantic
content in ontologies (i.e., the subclass hierarchy) can be used to
generate ``semantically closed'' feature vectors
\cite{Sokolov2013}. Alternatively, the output of semantic similarity
measures is widely used as features for machine learning
applications, for example in drug repurposing systems
\cite{Gottlieb2011} or identification of causative genomic variants
\cite{Robinson2013, pvp-main}.  Both of these approaches have in
common that the features generated through them contain no {\em
  explicit} information about the structure of the ontology and
therefore of the dependencies between the different features; these
dependencies are therefore no longer available as features for a
machine learning algorithm. In the case of semantic similarity
measures, the information in the ontology is used to define the
similarity but the information used to define the similarity is
subsequently reduced to a single point (the similarity value); in the
case of binary feature vectors, the ontology structure is used to
generate the values of the feature vector but is subsequently no
longer present or available to a machine learning algorithm. Feature
vectors that explicitly encode for {\em both} the ontology structure and
an entity's annotations would contain more information than either
information alone and may perform significantly better in machine
learning applications than alternative approaches.

Finally, semantic similarity measures are generally hand-crafted,
i.e., they are designed by an expert based on a set of assumptions
about how an ontology is used and what should constitute a
similarity. However, depending on the application of semantic
similarity, different features may be more or less relevant to define
the notion of similarity. It has previously been observed that
different similarity measures perform well on some datasets and tasks,
and worse on others \cite{Lord2003, Pesquita2008, Couto2009,
  Kulmanov2017}, without any measure showing clear superiority across
multiple tasks. One possible way to define a common similarity measure
that performs equally well on multiple tasks may be to establish a way
to {\em train} a semantic similarity measure in a data-driven
way. While this is not always possible due to the absence of training
data, when a set of desired outcomes (i.e., labelled data points) are
available, such an approach may result in better and more intuitive
similarity measures than hand-crafted ones.

We develop Onto2Vec, a novel method to jointly produce dense vector
representations of biological entities, their ontology-based
annotations, and the ontology structure used for annotations.  We
apply our method to the Gene Ontology (GO) \cite{GO} and generate
dense vector representations of proteins and their GO annotations. We
demonstrate that Onto2Vec generates vectors that can outperform
traditional semantic similarity measures in the task of
similarity-based prediction of protein-protein interactions; we also
show how to use Onto2Vec to train a semantic similarity measure in a
data-driven way, and use this to predict protein-protein interactions
and distinguish between the types of interactions. We further apply
Onto2Vec-generated vectors to clustering and show that the generated
clusters reproduce Enzyme Commission numbers of proteins.  The
Onto2Vec method is generic and can be applied to any set of entities
and their ontology-based annotations, and we make our implementation
freely available at
\url{https://github.com/bio-ontology-research-group/onto2vec}.
\section{Results}
\subsection{Onto2Vec}
We developed Onto2Vec, a method to learn dense, vector-based
representations of classes in ontologies, and the biological entities
annotated with classes from ontologies.  To generate the vector
representations, we combined symbolic inference (i.e., automated
reasoning) and statistical representation learning.  We first generated
vector-based representations of the classes in an ontology, and 
then extended our result to generate representations of biological
entities annotated with these classes.  The vector-based
representations generated by Onto2Vec provide the foundation for
machine learning and data analytics applications, including semantic
similarity applications.


Our main contribution with Onto2Vec is a method to learn a
representation of individual classes (and other entities) in an
ontology, taking into account all the axioms in an ontology that may
contribute to the semantics of a class, either directly or indirectly.
Onto2Vec uses an ontology $O$ in the OWL format, and applies the HermiT
OWL reasoner \cite{hermit} to infer new logical axioms (i.e.,
equivalent class axioms, subclass axioms, and disjointness axioms)(More technical details on the automated reasoning can be found in Section 5.2). We
call the union of the set of axioms in $O$ and the set of inferred
axioms the deductive closure of $O$, designated $O^\vdash$. In
contrast to treating ontologies as taxonomies or graph-based
structures \cite{rodriguez2018inferring}, we assume that every axiom in
$O$ (and consequently in $O^\vdash$) constitutes a sentence, and the
set of axiom in $O$ (and $O^\vdash$) a corpus of sentences. The
vocabulary of this corpus consists of the classes and relations that
occur in $O$ as well as the keywords used to formulate the OWL axioms
\cite{owl2, Grau2008}.  Onto2Vec then uses a skip-gram model to learn
a representation of each word that occurs in the corpus. The
representation of a word in the vocabulary (and therefore of a class
or property in $O$) is a vector that is predictive of words occurring
within a context window \cite{word2vec1, word2vec2}(More technical details on the representation learning can be found in Section 5.2).

Onto2Vec can also be used to learn vector-based representations of
biological entities that use ontologies for annotation and combine
information about the entities' annotations as well as the semantics of the
classes used in the annotation in a single representation. Trivially,
since Onto2Vec can generate representations of single classes in an
ontology, an entity annotated with $n$ classes, $C_1,...,C_n$, can be
represented as a (linear) combination of the vector representations of
these classes. For example, if an entity $e$ is annotated with $C_1$
and $C_2$, and $\nu(C_1)$ and $\nu(C_2)$ are the representations of
$C_1$ and $C_2$ generated through Onto2Vec, we can use
$\nu(C_1) + \nu(C_2)$ as a representation of $e$.  Alternatively, we
can use Onto2Vec directly to generate a representation of $e$ by
extending the axioms in $O$ with additional axioms that explicitly
capture the semantics of the annotation. If $O'$ is the ontology
generated from annotations of $e$ by adding new axioms capturing the
semantics of the annotation relation to $O$, then $e$ is a new class
or instance in $O'$ for which Onto2Vec will generate a representation
(since $e$ will become a word in the corpus of axioms generated from
$O'^\vdash$).

\begin{figure*}
  \centering
	\includegraphics[width=.9\textwidth]{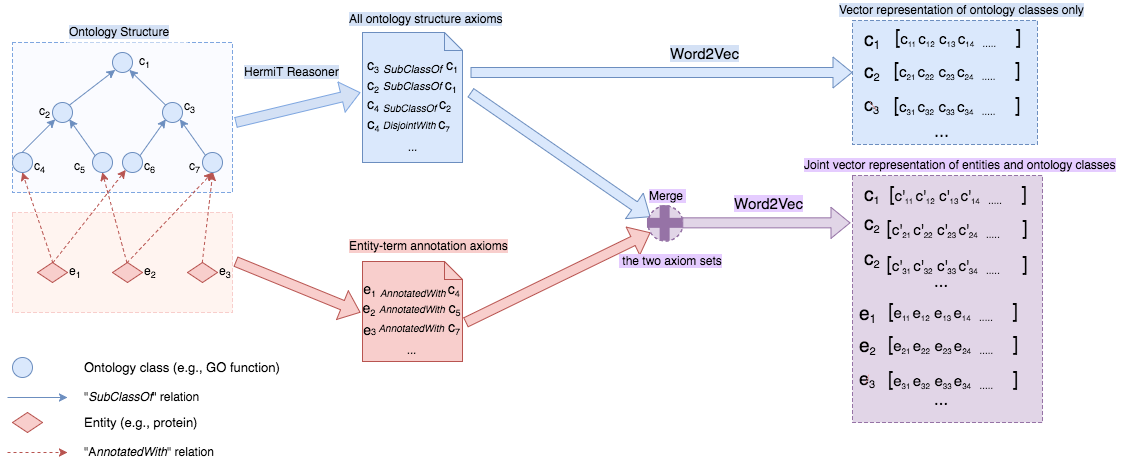}
	\caption{Onto2Vec Workflow. The blue-shaded part illustrates
          the steps to obtain vector representation for classes from
          the ontology. The purple-shaded part shows the steps to
          obtain vector representations of ontology classes and
          the entities annotated to these classes. }
	\label{fig:workflow}
\end{figure*}

As comprehensive use case, we applied our method to the GO, and to a
joint knowledge base consisting of GO and proteins with manual GO
annotations obtained from the UniProt database. To generate the latter
knowledge base, we added proteins as new entities and connected them using
a {\tt has-function} relation to their functions. We then applied
Onto2Vec to generate vector representations for each class in GO
(using a corpus based only on the axioms in GO), and further generate
joint representations of proteins and GO classes (using a corpus based
on the axioms in GO and proteins, and their annotations).  We further
generated protein representations by combining (i.e., adding) the GO
class vectors of the proteins' GO annotations (i.e., if a protein $p$
is annotated to $C_1,...,C_n$ and $\nu(C_1),...,\nu(C_2)$ are the
Onto2Vec-vectors generated for $C_1,...,C_n$, we define the
representation $\nu(p)$ of $p$ as
$\nu(p) = \nu(C_1) + ... + \nu(C_n)$).  In total, we generated 556,388
vectors representing proteins (each protein is represented three
times, either as a set of GO class vectors, the sum of GO class
vectors, or a vector jointly generated from representing {\tt
  has-function} relations in our knowledge base), and 43,828 vectors
representing GO classes. Figure \ref{fig:workflow} illustrates the
main Onto2Vec workflow to construct ontology-based vector
representations of classes and entities.
\subsection{Similarity-based prediction of biological relations}
\begin{table}[!h]
        {\begin{tabularx}{\columnwidth}{XXXX}\toprule  &
			{}& Yeast & Human\\\midrule
			$Resnik$&{}& 0.7942 & 0.7891 \\
			$Onto2Vec$& {}& 0.7701& 0.7614 \\
			$Onto2Vec\_NoReasoner$& {}& 0.7439& 0.7385 \\
			$Binary\_GO$ & {}& 0.6912& 0.6712 \\
			$Onto$\_$BMA$ &{}& 0.6741  & 0.6470\\
	     	$Onto$\_$AddVec$ &{}& 0.7139  & 0.7093\\
			$Onto2Vec\_LR$ &{}& 0.7959  & 0.7785\\
			$Onto2Vec\_SVM$ &{}& 0.8586 &0.8621\\
			$Onto2Vec\_NN$ &{}& \textbf{0.8869}  & \textbf{0.8931}\\
			$Binary\_GO\_LR$ &{}& 0.7009  & 0.7785\\
			$Binary\_GO\_SVM$ &{}& 0.8253  & 0.8068\\
			$Binary\_GO\_NN$ &{}& 0.7662  &0.7064\\ 
         \end{tabularx}}{}
       \caption{\label{tab:rocauc}AUC values of ROC curves for PPI
         prediction. The best AUC value among all methods is shown in
         bold.  {\tt Resnik} is a semantic similarity measure; {\tt
           Onto2Vec} is our method in which protein and ontology class
         representations are learned jointly from a single
         knowledgebase which is deductively closed; {\tt
           Onto2Vec\_NoReasoner} is identical to {\tt Onto2Vec} but
         does not use the deductive closure of the knowledge base;
         {\tt Binary\_GO} represents a protein's GO annotations as a
         binary vector (closed against the GO structure); {\tt
           Onto\_BMA} only generates vector representations for GO
         classes and compares proteins by comparing their GO
         annotations individually using cosine similarity and
         averaging individual values using the Best Match Average
         approach; {\tt Onto\_AddVec} sums GO class vectors to
         represent a protein. The methods with suffix {\tt LR}, {\tt
           SVM}, and {\tt NN} use logistic regression, a support
         vector machine, and an artificial neural network,
         respectively, either on the {\tt Onto2Vec} or the {\tt
           Binary\_GO} protein representations.}
\end{table}
We applied the vectors generated for proteins and GO classes to the
prediction of protein-protein interactions by functional, semantic
similarity.  
As a first experiment, we evaluated the accuracy of Onto2Vec in
predicting protein-protein interactions. For this purpose, we
generated several representations of proteins: first, we used Onto2Vec
to learn representations of proteins jointly with representations of
GO classes by adding proteins and their annotations to the GO using
the {\tt has-function} relations; second, we represented proteins as
the sum of the vectors representing the classes to which they are
annotated; and third, we represented proteins as the set of classes to
which they are annotated.

\begin{figure}
	\centering
	\begin{subfigure}{.5\textwidth}
		\centering
		\includegraphics[width=\linewidth, trim=4 4 4 4,clip]{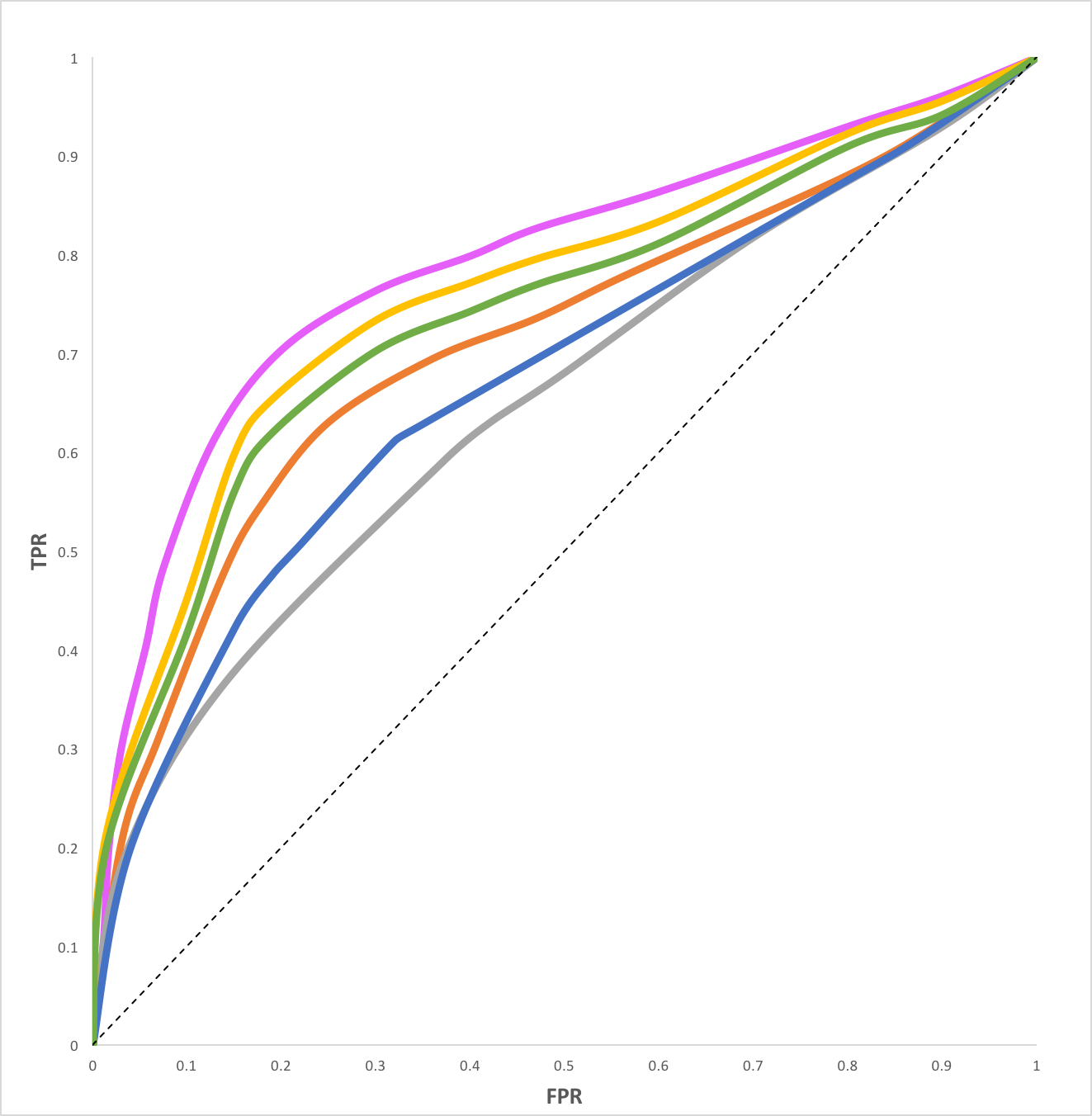}
		\caption{Human}
		\label{fig:sub1}
	\end{subfigure}%
	\begin{subfigure}{.5\textwidth}
		\centering
		\includegraphics[width=\linewidth,trim=4 4 4 4,clip,]{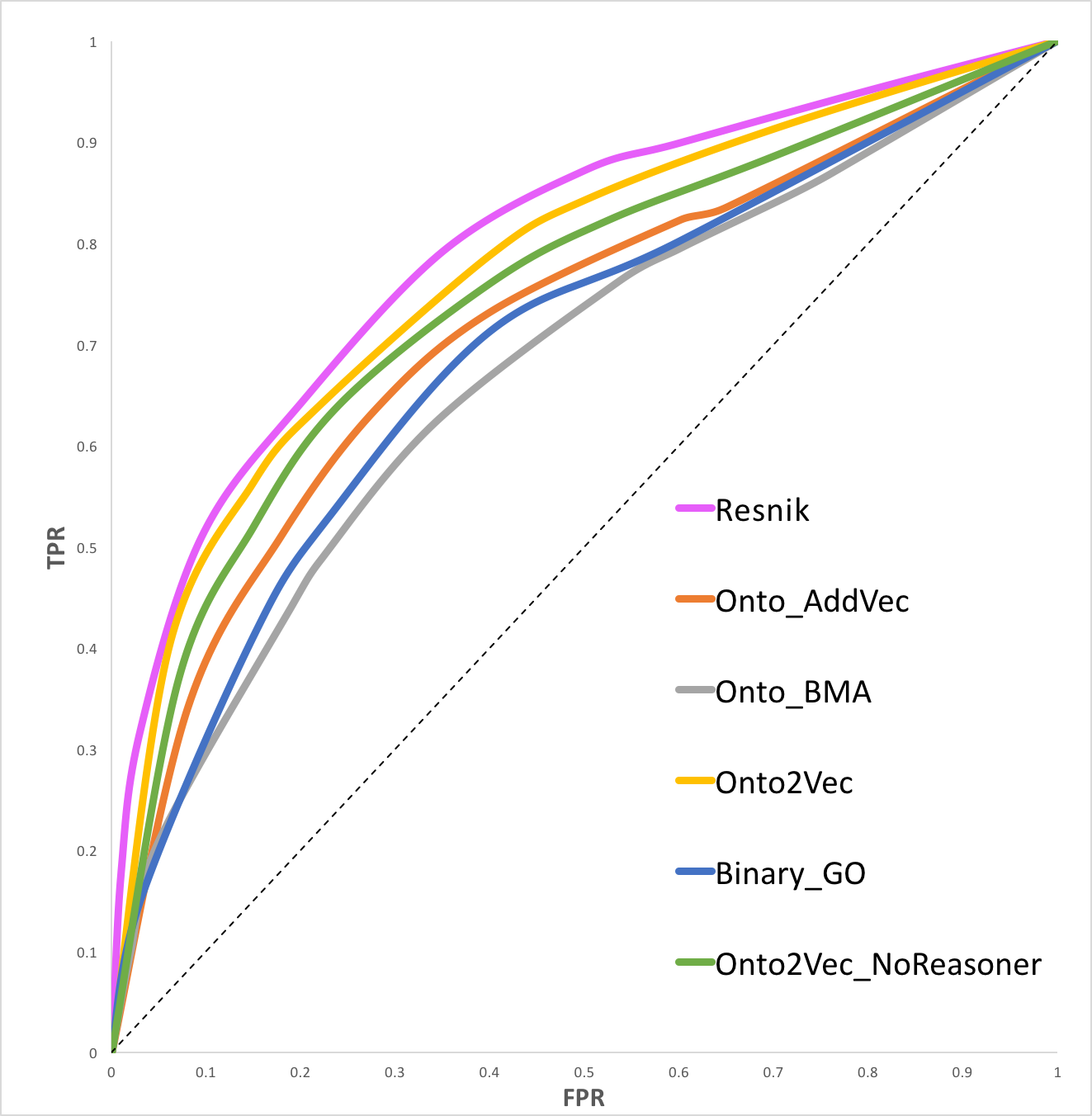}
		\caption{Yeast}
		\label{fig:sub2}
	\end{subfigure}
	\caption{\label{fig:unsupervised}ROC curves for PPI prediction for the
          unsupervised learning methods}
\end{figure}

We used cosine similarity to determine the similarity between
vectors. To compare sets of vectors (representing GO classes) to each
other, we used the Best Match Average (BMA) approach \cite{Couto2009},
where pairs of vectors are compared using cosine similarity.  We term
the approach in which we compared vectors generated from adding
proteins to our knowledge base {\em Onto2Vec}; {\em Onto\_AddVec} when
using cosine similarity between protein vectors generated by adding
the vectors of the GO classes to which they annotated; and {\em
  Onto\_BMA} when using the BMA approach to compare sets of GO
classes. To compare the different approaches for using Onto2Vec to the
established baseline methods, we further applied the Resnik's semantic
similarity measure \cite{resnik} with the BMA approach, and we
generated sparse binary vector representations from proteins' GO
annotations \cite{Sokolov2013} and compared them using cosine
similarity (termed {\em Binary\_GO}) (more technical details on the similarity measures used in Section 5.4).
Furthermore, to evaluate the contribution of using an automated
reasoner to infer axioms, we also included the results of using the
Onto2Vec approach without applying a reasoner.
We evaluated the performance of our method using protein-protein
interaction datasets in two species, human ({\em H. sapiens}) and
baker's yeast ({\em S. cerevisiae}).  Figure \ref{fig:unsupervised}
shows the ROC curves obtained for each approach on the human and the
yeast datasets; the area under the ROC curve (ROCAUC) values are shown
in Table \ref{tab:rocauc}. We found that Resnik's semantic similarity
measure performs better than all other methods we evaluated, and that
the Onto2Vec representation based on generating representations
jointly from proteins and GO classes performs second best.  These
results demonstrate that Resnik's semantic similarity measure, which
determines similarity based on the information content of ontology
classes as well as the ontology structure, is better suited for this
application than our Onto2Vec representations using cosine similarity.

However, a key feature of Onto2Vec representations is their ability to
encode for annotations and the ontology structure; while cosine
similarity (and the derived measures) can determine whether two
proteins are similar, certain classes and axioms may contribute more
to predicting protein-protein interactions than others. To test
whether we can use the information in Onto2Vec representations in such
a way, we used supervised machine learning to train a similarity
measure that is predictive of protein-protein interactions. To this
end we used three different machine learning methods, logistic
regression, support vector machines (SVMs), and neural networks (more technical details available in Section 5.5). To
obtain a baseline comparison, we also trained each model using the
Binary\_GO protein representations.

\begin{figure}
	\centering
	\begin{subfigure}{.5\textwidth}
		\centering
		\includegraphics[width=\linewidth, trim=4 4 4 4,clip]{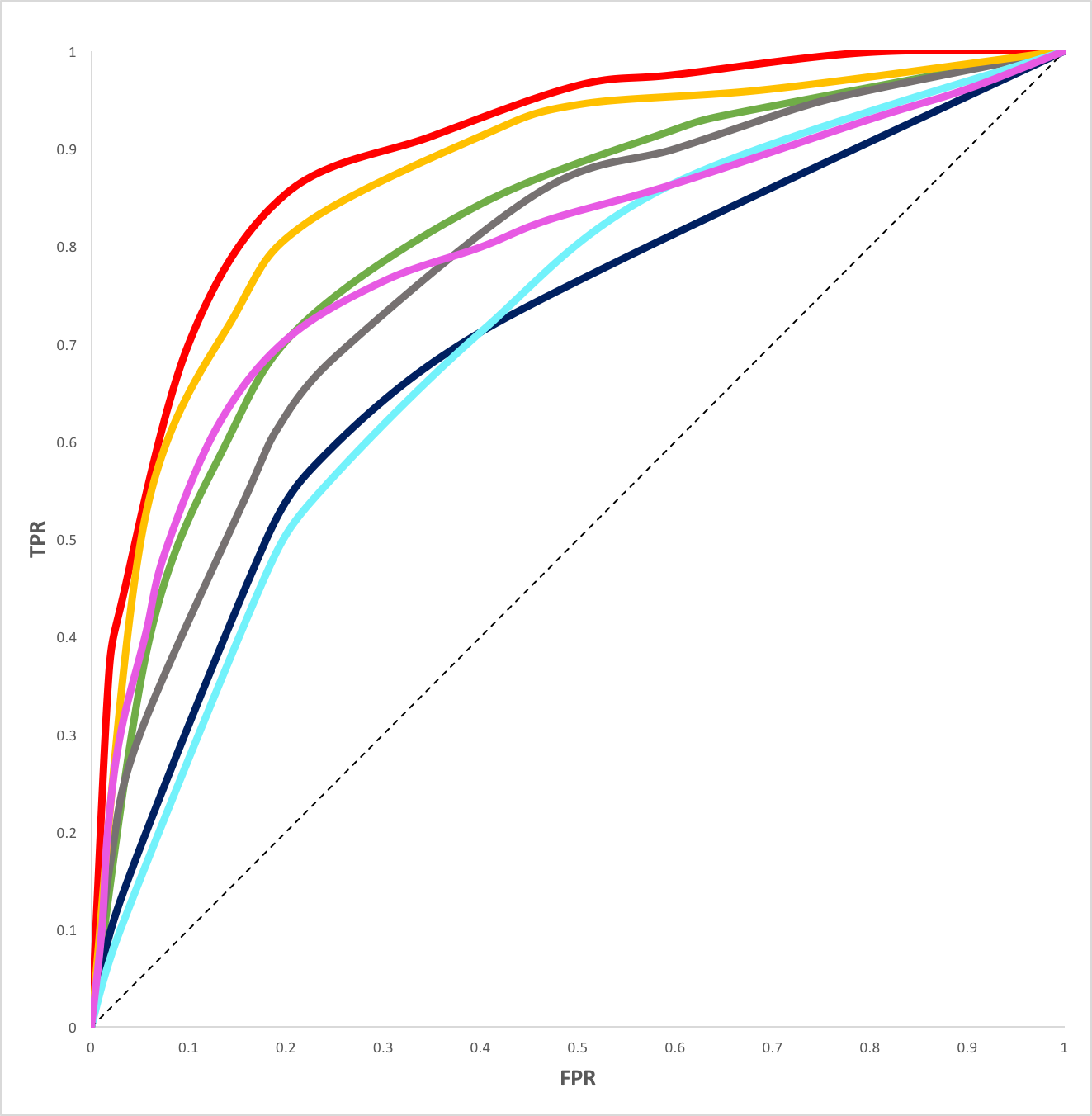}
		\caption{Human}
		\label{fig:sub1}
	\end{subfigure}%
	\begin{subfigure}{.5\textwidth}
		\centering
		\includegraphics[width=\linewidth,trim=4 4 4 4,clip,]{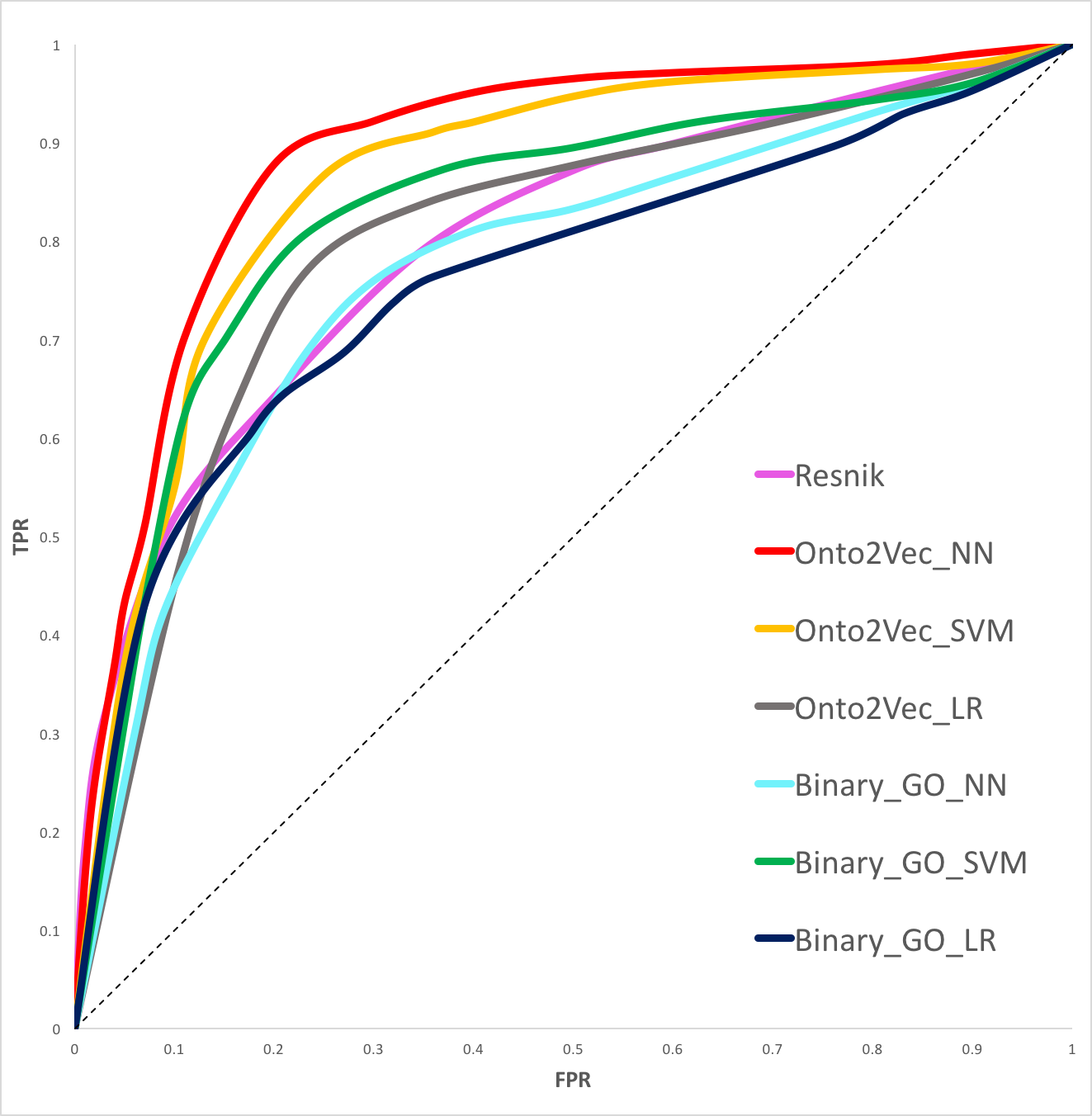}
		\caption{Yeast}
		\label{fig:sub2}
	\end{subfigure}
	\caption{ROC curves for PPI prediction for the supervised learning methods in addition to Resnik measure}
	\label{fig:supervised}
\end{figure}

Each model uses a pair of protein vectors as inputs and is trained
to predict whether the proteins provided as input interact or
not. Each supervised model also outputs intermediate confidence values
and can therefore be considered to output a form of similarity.
The ROC curves of all trained models using the Onto2Vec and binary
representations of proteins are shown in Figure \ref{fig:supervised},
and their ROCAUC values are reported in Table \ref{tab:rocauc}. We
observed that the supervised models (i.e., the ``trained'' semantic
similarity measures) using Onto2Vec protein representations outperform
the use of pre-defined similarity measures in all experiments; while
logistic regression performs comparable to Resnik semantic similarity,
both SVMs and artificial neural networks can learn similarity measures
that predict protein-protein interactions significantly better than
any pre-defined similarity measure. Onto2Vec representations further
outperform the sparse binary representations of protein functions,
indicating that the combination of annotations and ontology axioms indeed
results in improved predictive performance.
We further tested whether the supervised models (i.e., the trained
semantic similarity measures) can be used as similarity measures so
that higher similarity values represent more confidence in the
existence of an interaction. We used the confidence scores associated
with protein-protein interactions in the STRING database and determined
the correlation between the prediction score of our trained models and
the confidence score in STRING. Table \ref{tab:correlation} summarizes
the Spearman correlation coefficients for each of the methods we
evaluated. We found that our trained similarity measures correlate more
strongly with the confidence measures provided by STRING than other
methods, thereby providing further evidence that Onto2Vec
representations encode useful information that is predictive of
protein-protein interactions.
\begin{table}[!h]
  {\begin{tabularx}{\columnwidth}{XXXX}\toprule & {}& Yeast &
      Human\\\midrule
     $Resnik$&{}&  0.1107 & 0.1151 \\
     $Onto2Vec$& {}& 0.1067 & 0.1099\\
     $Binary\_GO$ & {}& 0.1021&0.1031 \\
     $Onto2Vec\_LR$ &{}& 0.1424  & 0.1453\\
     $Onto2Vec\_SVM$ &{}& 0.2245  & 0.2621\\
     $Onto2Vec\_NN$ &{}& \textbf{0.2516}  & \textbf{0.2951}\\
     $Binary\_GO\_LR$ &{}& 0.1121  & 0.1208\\
     $Binary\_GO\_SVM$ &{}& 0.1363  & 0.1592\\
     $Binary\_GO\_NN$ &{}& 0.1243 & 0.1616\\
	\end{tabularx}}{}
      \caption{Spearman correlation coefficients between STRING
        confidence scores and PPI prediction scores of different
        prediction methods. The highest absolute correlation across
        all methods is highlighted in bold.  \label{tab:correlation}}
\end{table}
Finally, we trained our models to separate protein-protein interactions
into different interaction types, as classified by the STRING
database: {\em reaction}, {\em activation}, {\em binding}, and {\em
  catalysis}. For comparison, we also reported results when using sparse
binary representations of proteins in the supervised models, and we
reported Resnik semantic similarity and Onto2Vec similarity results
(using cosine similarity). Table \ref{tab:mechanism} summarizes the
results. While Resnik semantic similarity and Onto2Vec similarity
cannot distinguish between different types of interaction, we find
that the supervised models, in particular the multiclass SVM and
artificial neural network, are capable when using Onto2Vec vector
representations to distinguish between different types of interaction.
In addition, the Onto2Vec representations perform better than sparse
binary vectors, indicating further that encoding parts of the ontology
structure can improve predictive performance.
\begin{table*}[!htbp]
	\tiny
	\centering
	\begin{tabularx}{\textwidth}{XX|XXXX|XXXXX}
		\toprule
		{} & {} &  \multicolumn{4}{c}{\textbf{Yeast}} & \multicolumn{4}{c}{\textbf{Human}}\\
		\midrule
		{}   & {} & \textbf{Reaction} & \textbf{Activation} & \textbf{Binding} &\textbf{Catalysis}    &  \textbf{Reaction} & \textbf{Activation}  & \textbf{Binding} & \textbf{Catalysis}\\
		$Resnik$    &{}& 0.5811 & 0.6023 &0.5738  & 0.5792   & 0.5341 & 0.5331 & 0.5233 & 0.5810\\
		$Onto2Vec$    &{}& 0.5738 & 0.5988 & 0.5611 & 0.5814   & 0.5153 & 0.5104 & 0.5073 & 0.6012\\
		$Onto2Vec\_LR$    &{}&   0.7103 & 0.7011 & 0.6819 & 0.6912   & 0.7091 & 0.6951 & 0.6722 & 0.6853\\
		$Onto2Vec\_multiSVM$   &{} &  \textbf{0.7462} & \textbf{0.7746} & 0.7311 & \textbf{0.7911}  & \textbf{0.7351} & \textbf{0.7583} & 0.7117 & \textbf{0.7724}\\
		$Onto2Vec\_NN$   &{}&   0.7419& 0.7737 & \textbf{0.7423} & 0.7811   & 0.7265 & 0.7568 & \textbf{0.7397} & 0.7713\\
		$Binary\_GO\_LR$   & {} &  0.6874& 0.6611 &0.6214 & 0.6433  & 0.6151 & 0.6533 & 0.6018 & 0.6189\\
		$Binary\_GO\_multiSVM$   & {} &  0.7455& 0.7346& 0.7173 & 0.7738  & 0.7246 & 0.7132 & 0.6821& 0.7422\\
		$Binary\_GO\_NN$   & {} &  0.7131  & 0.6934& 0.6741  & 0.6838   & 0.6895  & 0.6803&  0.6431& 0.6752\\
		
	\end{tabularx}
	\caption{AUC values of the ROC curves for PPI interaction type
          prediction. The best AUC value for each action is shown in
          bold.\label{tab:mechanism}}
\end{table*}
\subsection{Clustering and visualization}
\begin{figure*}
  \centering
	\includegraphics[width=.9\textwidth]{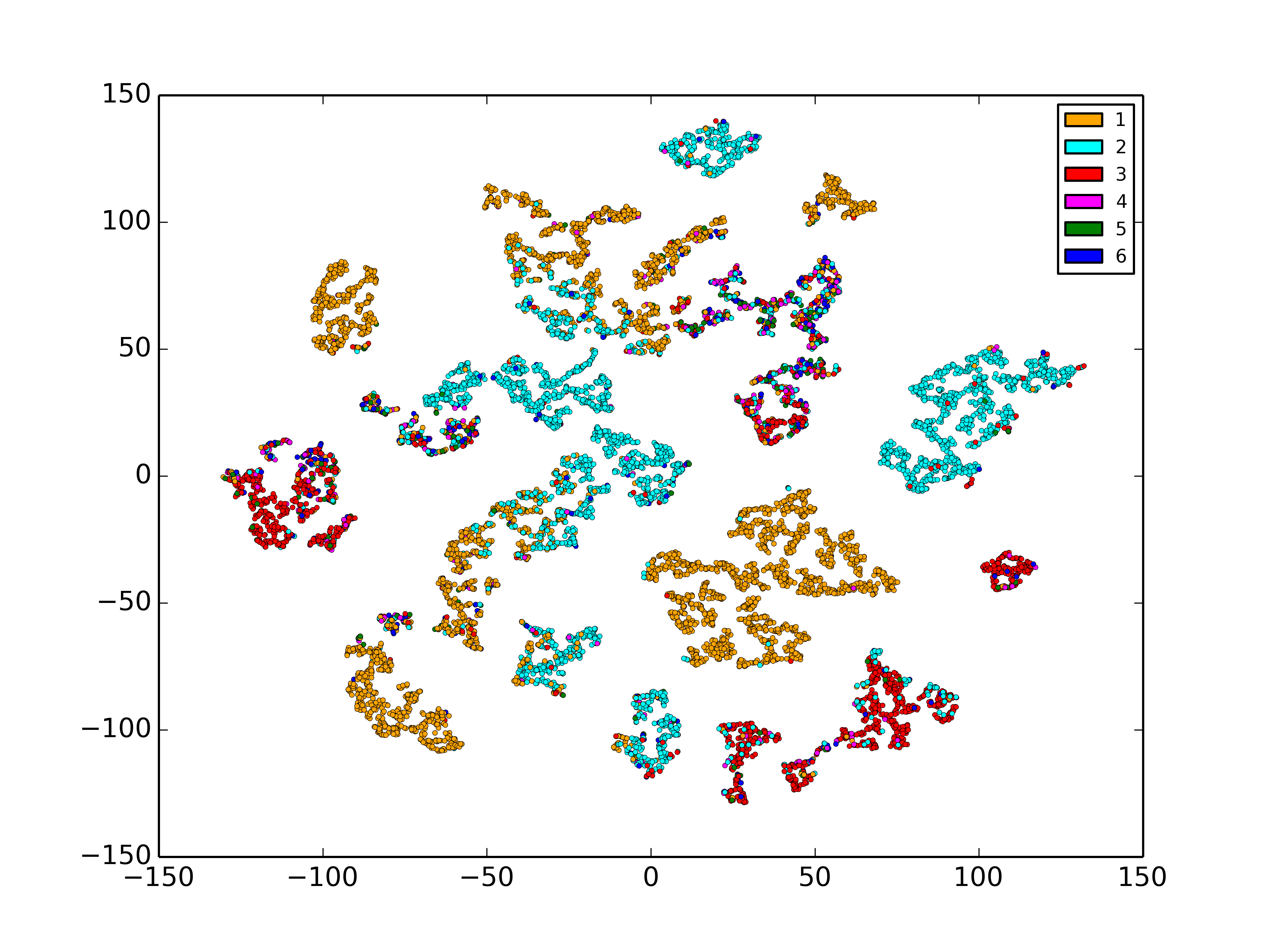}
	\caption{t-SNE visualization of 10,000 enzyme vectors color-coded by their first level EC category (1,2,3,4,5 or 6).}
	\label{fig:tsne}
\end{figure*}

Onto2Vec representations can not only be used to compute semantic
similarity or form part of supervised models, but can also provide the
foundation for visualization and unsupervised clustering.  The ability
to identify sets of biological entities which are more similar to each
other within a dataset can be used for clustering and identifying groups
of related biological entities.  We visualized the GO-based vector
representations of proteins generated by Onto2Vec. Since the Onto2Vec
representations are of a high dimensionality, we applied the t-SNE
dimensionality reduction \cite{tsne} to the vectors and represented
10,000 randomly chosen enzyme proteins in Figure \ref{fig:tsne} (We refer the readers to section 5.7 for more technical details in t-SNE ).

The visual representation of the enzymes shows that the proteins are
separated and form different functional groups. To explore what kind
of information these groups represent, we identified the EC number for each enzyme and colored the enzymes
in six different groups depending on their top-level EC category. We
found that some of the groups that are visually separable represent
mainly enzymes within a single EC top-level category. To quantify
whether Onto2Vec similarity is representative of EC categorization, we
applied $k$-means clustering ($k=6$) to the protein representations. We
evaluated cluster purity with respect to EC top-level classification
and found that the purity is $0.42$; when grouping enzymes based on
their second-level EC classification ($k=62$), cluster purity
increases to $0.60$.
\section{Discussion}
\subsection{Ontologies as graphs and axioms}
We have developed Onto2Vec, a novel method for learning feature
vectors for entities in ontologies.  There have been several recent
related efforts that use unsupervised learning to generate dense
feature vectors for structured and semantically represented
data. Notably, there is a large amount of work on knowledge graph
embeddings \cite{transe, Nickel2016, Perozzi2014}, i.e., a set of
feature learning methods applicable to nodes in heterogeneous graphs,
such as those defined by Linked Data \cite{Bizer2009}. These methods
can be applied to predict new relations between entities in a
knowledge graph, perform similarity-based predictions, reason by
analogy, or in clustering \cite{Nickel2016review}. However, while some
parts of ontologies, such as their underlying taxonomy or partonomy,
can naturally be expressed as graphs in which edges represent
well-defined axiom patterns \cite{Smith2005, h20}, it is challenging
to represent the full semantic content of ontologies in such a way
\cite{rodriguez2018inferring}.

It is possible to materialize the implicit, inferred content of
formally represented knowledge bases through automated reasoning, and
there is a long history in applying machine learning methods to the
deductive closure of a formalized knowledge base \cite{Valiant1985,
  Bergadano1991}.  Similar approaches have also been applied to
knowledge graphs that contain references to classes in ontologies
\cite{alsharani17}. However, these approaches are still limited to
representing only the axioms that have a materialization in a
graph-based format. Onto2Vec is, to the best of our knowledge, the
first approach which applies feature learning to arbitrary OWL axioms
in biomedical ontologies and includes a way to incorporate an
ontology's deductive closure in the feature learning process. While
Onto2Vec can be used to learn feature representations from
graph-structures (by representing graph edges as axioms, or triples),
the opposite direction is not true; in particular axioms involving
complex class expressions, axioms involving disjointness, and axioms
involving object property restrictions, are naturally included by
Onto2Vec while they are mostly ignored in feature learning methods
that rely on graphs alone.


\subsection{Towards ``trainable'' semantic similarity measures}

Another related area of research is the use of semantic similarity
measures in biology.  Onto2Vec generates feature representations of
ontology classes, or entities annotated with several ontology classes,
and we demonstrate how to use vector similarity as a measure of
semantic similarity. In our experiments, we were able to almost match
the performance of an established semantic similarity measure
\cite{resnik} when using cosine similarity to compare proteins.  It is
traditionally challenging to evaluate semantic similarity measures,
and their performances differ between biological problems and datasets
\cite{Couto2009, Lord2003, Pesquita2008, Kulmanov2017}.  The main
advantage of Onto2Vec representations is their ability to be used in
trainable similarity measures, i.e., problem- and dataset-specific
similarity measures generated in a supervised way from the available
data. The training overcomes a key limitation in manually created
semantic similarity measures: the inability to judge {\em a priori}
how each class and relation (i.e., axiom) should contribute to
determining similarity. For example, for predicting protein-protein
interactions, it should be more relevant that two proteins are active
in the same (or neighboring) cellular component than that they both
have the ability to regulate other proteins. Trainable similarity
measures, such as those based on Onto2Vec, can identify the importance
of certain classes (and combinations of classes) with regard to a
particular predictive task and therefore improve predictive
performance significantly.

Furthermore, Onto2Vec does not only determine how classes, or their
combinations, should be weighted in a similarity computation. Semantic
similarity measures use an ontology as background knowledge to
determine the similarity between two (sets of) classes; how the
ontology is used is pre-determined and constitutes the main
distinguishing feature among semantic similarity measures
\cite{Couto2009}. Since Onto2Vec vectors represent both an entity's
annotations and (parts of) the ontology structure, the way in which
this structure is used to compute similarity can also be determined in
a data-driven way through the use of supervised learning; it may
even be different between certain branches of an ontology. We
demonstrate that supervised measures outperform binary
representations, which shows that combining ontology-based annotations
and the ontology structure in a single representation has clear
advantages.

\section{Conclusions}

Onto2Vec is a method that combines neural and symbolic methods in
biology, and demonstrates significant improvement over state-of-the-art methods. There is now an increasing interest in the integration of
neural and symbolic approaches to artificial intelligence
\cite{Besold2017}.  In biology and biomedicine, where a large amount
of symbolic structures (ontologies and knowledge graphs) are in use,
there are many potential applications for neural-symbolic systems
\cite{datascience}.

The current set of methods for knowledge-driven analysis (i.e.,
analysis methods that specifically incorporate symbolic structures and
their semantics) in biology is limited to ontology enrichment analysis
\cite{Subramanian2005}, applications of semantic similarity
\cite{Couto2009}, and, to a lesser degree, network-based approaches
\cite{Dutkowski2012}. With Onto2Vec, we introduce a new method in the
semantic analysis toolbox, specifically targeted at computational
biology and the analysis of datasets in which ontologies are used for
annotation. While we already demonstrate how Onto2Vec representations
can be used to improve predictive models for protein-protein
interactions, additional experiments with other ontologies will likely
identify more areas of applications.
We expect that future research on neural-symbolic systems will further
extend our results and enable more comprehensive analysis of symbolic
representations in biology and biomedicine.

\section{Materials and Methods} 
\subsection{Data set}
We downloaded the Gene Ontology (GO) in OWL format from the Gene
Ontology Consortium Website
(\url{http://www.geneontology.org/ontology/}) on 2017-09-13. We
obtained the GO protein annotations from the UniProt-GOA website
(\url{http://www.ebi.ac.uk/GOA}) on 2017-09-26. We filtered all
automatically assigned GO annotations (with evidence code {\tt IEA} as
well as {\tt ND}) which results in $5.5\times 10^6$ GO annotations.

We obtained the protein-protein interaction networks for both
yeast ({\em Saccharomyces cerevisiae}), and humans ({\em Homo
  sapiens}) from the STRING database \cite{string}
(\url{http://string-db.org/}) on 2017-09-16. The human protein dataset
contains 19,577 proteins and 11,353,057 interactions while the yeast
dataset contains 6,392 proteins and 2,007,135 interactions. We extracted
Enzyme Commission (EC) number annotations for 10,000 proteins from
Expasy \cite{expasy}
(\url{ftp://ftp.expasy.org/databases/enzyme/enzyme.dat}) on 2017-10-4.

\subsection{Automated reasoning}
We used the OWL API version 4.2.6 \cite{owl2007igniting} to process
the GO in OWL format \cite{gene2013gene}. Our version of GO contains
577,454 logical axioms and 43,828 classes. We used the HermiT reasoner
(version 1.3.8.413) \cite{hermit} to infer new logical axioms from the
asserted ones. We used HermiT as it supports all OWL 2 DL axioms and
has been optimized for large ontologies \cite{hermit}. These
optimizations make HermiT relatively fast which is particularly
helpful when dealing with ontologies of the size of GO. We infer three
types of axioms: subsumption, equivalence and disjointness, resulting
in 80,133 new logical axioms that are implied by GO's axioms and
materialized through HermiT.

\subsection{Representation learning using Word2Vec}
We treated an ontology as a set of axioms, each of which constitutes a
sentence. To process the axioms syntactically, we used the Word2Vec
\cite{word2vec1, word2vec2} methods.  Word2Vec is a set of
neural-network based tools which generate vector representations of
words from large corpora. The vector representations are obtained in
such a way that words with similar contexts tend to be close to each
other in the vector space. 

Word2Vec can use two distinct models: the continuous bag of word
(CBOW), which uses a context to predict a target word, and the
skip-gram model which tries to maximize the classification of a word
based on another word from the same sentence. The main advantage of
the CBOW model is that it smooths over a lot of the distributional
information by treating an entire context as one observation, while
the skip-gram model treats each context-target as a new observation,
which works better for larger datasets. The skip-gram model has the
added advantage of producing higher quality representation of rare
words in the corpus \cite{word2vec1,word2vec2}. Here, we chose the
skip-gram architecture since it meets our need to produce high quality
representations of all biological entities occurring in our large
corpus, including infrequent ones. Formally, given a sequence of
training words $\omega_1$, $\omega_2$, ..., $\omega_T$, the skip-gram
model aims to maximize the following average log likelihood:
\begin{equation}
\frac{1}{T}\sum_{t=1}^{T}\sum_{-c\leq j\leq c, j\neq0}log\,p(\omega_{t+j}|\omega_t),\label{eq:01}
\end{equation}
where $c$ is the size of the training context, $T$ is the size of the
set of the training words and $\omega_i$ is the $i$-th training word
in the sequence. We identified an optimal set of parameters of the
skip-gram model through limited gridsearch on the following
parameters: the size of the output vectors on the interval [50--250] using a step size of 50,
the number of iterations on the interval [3-5] and negative sampling
on the interval [2-5] using a step size of 1. Table~1\vphantom{\ref{Tab:01}}
shows the parameter values we used for the skip-gram in our work.
\begin{table}[!h]
\center
{\begin{tabular}{@{}lp{7cm}ll@{}}\toprule Parameter &
			Definition & Default value\\\midrule
			\vbox{\hbox{\strut $sg$}}& \vbox{\hbox{\strut Choice of training algorithm}\hbox{\strut $sg$= 1 skip-gram}\hbox{\strut $sg$= 0 CBOW}} & 1 \\
			$size$ & Dimension of the obtained vectors & 200 \\
			$min$\_$count$ &Words with frequency lower than this value will be ignored  & 1\\
			$ window$&Maximum distance between the current and the predicted word & 10\\
			$iter$&Number of iterations&5\\
			$negative$ & Whether negative sampling will be used and how many ``noise words" would be drawn& 4\\
	\end{tabular}}{}
      \caption{Parameter we use for training the Word2Vec model.\label{Tab:01}} 
    \end{table}
\subsection{Similarity}
We used cosine similarity to determine similarity between feature
vectors generated by Onto2Vec. The cosine similarity, $cos_{sim}$, between
two vectors $A$ and $B$ is calculated as
\begin{equation}
cos_{sim}(A,B)=\frac{A\cdot B}{ ||A||  ||B||},\label{eq:02}
\end{equation}
where $A \cdot B$ is the dot product of $A$ and  $B$.

We used Resnik's semantic similarity measure \cite{resnik} as the baseline
for comparison. Resnik's semantic similarity measure is widely used in
biology \cite{Couto2009}. It is based on the notion of
information content which quantifies the specificity of a given class
in the ontology. The information content of a class $c$ is defined as
the negative log likelihood, $- \log p(c)$, where $p(c)$ is the
probability of encountering an instance or annotation of class
$c$. Given this definition of information content, Resnik similarity
is formally defined as:
\begin{equation}
sim(c_1,c_2)=- \log p(c_{MICA}), \label{eq:03}
\end{equation}
where $c_{MICA}$ is the most informative common ancestor of $c_1$ and
$c_2$ in the ontology hierarchy, defined as the common ancestor of
$c_1$ and $c_2$ with the highest information content value. 
Resnik's similarity measure only measures the similarity between two
ontology classes. We applied the Best Match Average (BMA) method
\cite{bma} to compute the similarity between two sets of classes.
For two biological entities $e_1$ and $e_2$, the BMA is defined as:
\begin{equation}
\resizebox{.9\hsize}{!}{$BMA(e_1,e_2)=\frac{1}{2}\bigg(\frac{1}{n}\sum_{c_1\in
    S_1}\max_{c_2\in S_2}sim(c_1,c_2)+\frac{1}{m}\sum_{c_2\in
    S_2}\max_{c_1\in S_1}sim(c_1,c_2)\bigg)$},\label{eq:04} 
\end{equation}
\noindent where $S_1$ is the set of ontology concepts that $e_1$ is annotated
with, $S_2$ is the set of concepts that $e_2$ is annotated with, and
$sim(c_1,c_2)$ is the similarity value between concept $c_1$ and
concept $c_2$, which could have been calculated using Resnik
similarity or any other semantic similarity measure (e.g., cosine
similarity).

\subsection{Supervised Learning}
We used supervised learning to train a similarity measure between two
entities that is predictive of protein-protein interactions.
We applied our method to two datasets, one for protein--protein
interactions in yeast and another in human. We filtered the STRING
database and kept only proteins with experimental annotations which is
a total of 18,836 proteins in the human dataset and 6,390 proteins in
the yeast dataset. We randomly split each dataset into 70\% and 30\%
for training and testing respectively . The positive pairs are all
those reported in the STRING database, while the negative pairs are
randomly sub-sampled among all the pairs not occurring in STRING, in
such a way that the cardinality of the positive set and that of the
negative set are equal for both the testing and the training datasets.

We used logistic regression, support vector machines (SVMs), and
artificial neural networks (ANNs) to train a classifier for
protein-protein interactions. 
We trained each of these methods by providing a pair of proteins
(represented through their feature vectors) as input and predicting
whether the pair interacts or not. The output of each method varies
between $0$ and $1$, and we used the prediction output as a similarity
measure between the two inputs.

Logistic regression does not require any selection of parameters. We
used the SVM with a linear kernel and sequential minimal optimization.
Our ANN structure is a feed-forward network with four layers: the
first layer contains 400 input units; the second and third layers are
hidden layers which contain 800 and 200 neurons, respectively; and the
fourth layer contains one output neuron. We optimized parameters using
a limited manual search based on best practice guidelines
\cite{hunter2012selection}. We optimized the ANN using binary cross
entropy as the loss function.

In addition to binary classification, we also trained multi-class
classifiers to predict the type of interaction between two types of
proteins. We used a multi-class SVM as well as ANNs; the parameters we
used are identical to the binary classification case, except that we
used an ANN architecture with more than one output neuron (one for each
class).  We implemented all supervised learning methods in
MATLAB.



\subsection{Evaluation}
The Receiver Operating Characteristic (ROC) curve is a widely used
evaluation method to assess the performance of prediction and
classification models. It plots the true-positive rate (TPR or
sensitivity) defined as $TPR=\frac{TP}{TP+FN}$ against the
false-positive rate (FPR or $1-$specificity) defined as
$FPR=\frac{FP}{FP+TN}$, where $TP$ is the number of true positives,
$FP$ is the number of false positives and $TN$is the number of true
negatives \cite{Fawcett2006}.
We used ROC curves to evaluate protein-protein interaction prediction
of our method as well as baseline methods, and we reported the area
under the ROC curve (ROCAUC) as a quantitative measure of classifier
performance. In our evaluation, the $TP$ value is the number of
protein pairs occurring in STRING regardless of their STRING
confidence score and which have been predicted as interacting. The
$FP$ value is the number of protein pairs which have been predicted as
interacting but do not appear in STRING. And the
$TN$ is the number of protein pairs predicted as non-interacting and
which do not occur in the STRING database.



\subsection{Clustering and Visualization}
For visualizing the ontology vectors we generated, we used the t-SNE
\cite{tsne} method to reduce the dimensionality of the vectors to 2
dimensions, and plotted the vectors in the 2D space.  t-SNE is similar to
principal component analysis but uses probability distributions to
capture the non-linear structure of the data points, which linear
dimensionality reduction methods, such as PCA, cannot achieve
\cite{tsne}. We used a perplexity value of $30$ when applying t-SNE.

The k-means algorithm is used to cluster the protein vectors, and we quantitatively
measured the quality of these clusters with respect to EC families by
using cluster purity. Cluster purity is defined as:
\begin{equation}
purity(T,C)=\frac{1}{N}\sum_{i=0}^{k}\max_{j}(c_k\cup t_j),
\end{equation}
where $N$ is the total number of data points, $C={c_1,c_2,...,c_k}$ is
the set of clusters, and $T={t_1,t_2,...,t_J}$ is the set of classes
which is in this case the set of EC families. Since there are six first-level EC categories, the number of classes in this case
is six and the number of clusters used in k-means is also set to
six.

\section*{Funding}

The research reported in this publication was supported by the King Abdullah
University of Science and Technology (KAUST) Office of Sponsored Research
(OSR) under Award No. URF/1/1976-04, URF/1/1976-06, URF/1/3450-01-01 and URF/1/3454-01-01.

\nolinenumbers

\bibliography{library}

\bibliographystyle{abbrv}

\end{document}